\documentclass{article}

\usepackage{arxiv}

\usepackage[utf8]{inputenc} 
\usepackage[T1]{fontenc}    
\usepackage{hyperref}       
\usepackage{url}            
\usepackage{booktabs}       
\usepackage{amsfonts}       
\usepackage{nicefrac}       
\usepackage{microtype}      
\usepackage{cleveref}       
\usepackage{lipsum}         
\usepackage{graphicx}
\usepackage{natbib}
\usepackage{doi}

\DeclareMathSymbol{\blacktriangle} {\mathord}{AMSa}{"4E}
\DeclareMathSymbol{\blacksquare} {\mathord}{AMSa}{"04}
\DeclareMathSymbol{\blacklozenge} {\mathord}{AMSa}{"07}

\title{Use of Euler's theorem in the elucidation of economic concepts 
in goods exchange}

\newif\ifuniqueAffiliation

\uniqueAffiliationtrue

\ifuniqueAffiliation 
\author{ 
{
\hspace{1mm}Juan Villegas-Febres}\thanks{Also: juanvillegas.febres@gmail.com} \\
	Grupo de Qu\'imica Te\'orica QUIFFIS\\ 
	Facultad de Ciencias\\
	Universidad de Los Andes\\
	M\'erida 5101\\
	Venezuela\\
	\texttt{juancv@ula.ve} \\
}
\else

\usepackage{authblk}

\author[1]{

	\hspace{1mm}Juan C. Villegas-Febres\thanks{\texttt{juancv@ula.ve}}}
}

\affil[1]{Grupo de Qu\'imica Te\'orica: Quimicof\'isica de Fluidos y Fen\'omenos Interfaciales (QUIFFIS), Facultad de Ciencias, Universidad de Los Andes, M\'erida 5101, Venezuela}

\fi

\hypersetup{
pdftitle={Use of Euler's theorem in the elucidation of economic concepts 
in goods exchange},
pdfauthor={Juan C. Villegas-Febres},
pdfkeywords={Entropy, Economy, Price, Total Revenue, Indexing},
}

\begin{document}
\maketitle

\begin{abstract}
Starting from a plausible assumption about the Total Revenue concept, a system of economic agents, that simulates the exchange of goods, is studied. Following a methodology equivalent to that used in the statistical-mechanical determination of the distribution of energies in a physical system, it is shown that the price of the exchanged goods arises {\it naturally}, as well as the well-known ``Law of demand"\ . It is also shown that by using Euler's Theorem of homogeneous functions, the conditions emerge for the appearance of wholesale and retail prices. We found also that the Total Revenue is the product of two factors: the configurational entropy and a "free energy" type one. We study numerically the case in which the exchanged good is money itself, in systems in which economic agents have a "roof" on the amount of money they can possess. It is shown numerically that the increase in the quantity of money in the hands of the agents necessarily leads to its depreciation, and only the expansion of the economy, understood here as raising the "roof", minimizes this loss of purchasing power. The concept of {\it indexing} is discussed.
\end{abstract}


\section{Introduction}

The economy is a transversal issue that affects all areas of human life, regardless of the objective and subjective conditions of the historical moment in which we have had to live. As a discipline of study, economics works mostly with "laws"\ of an observational type, such as the {\it Law of demand}, or the {\it Law of diminishing returns}, fundamentally the product of experience accumulated over generations. Science, on the contrary, is endowed with very detailed laws, theorems, real and virtual experiments, the result of the application of the Scientific Method. One way to give economic studies a scientific basis is to approach them using, fundamentally, although not necessarily exclusively, thermodynamics and statistical mechanics, giving rise to what has been called {\it Econophysics}.

A recurring theme has been the distribution of money in a model society, as a way of understanding the mechanisms that underlie the unequal distribution of wealth and/or income, observed in real human societies. The simulation experiments are focused on the study of money exchange, observing, in general, a distribution of {\it Boltzmann type} for the economic agents that earn less income or wealth and the rest is {\it Pareto type} for the higher purchasing power strata. One aspect that has been explored, in order to seek improvements in socio-economic inequality, is the effect of limiting the maximum amount of money and/or wealth that an individual can possess (the so-called ``roof"\ in levels accessible to the system).

In these studies the money is worked, as is done analogously with energy in physical systems, under the principle of
the ``conservation of money"\
\footnote{That is, money is neither created nor destroyed in the system, what it does is transfer from one agent to another economic agent.},
and the effect of entropy is considered a side issue. In other words, the description of the economic system gravitates around a purely ``energetic"\ approach. Studies focused on entropy, however, are essentially conceptual, and do not culminate in practical applications. In 1948 E. Shannon ``discovered "\ the entropy studying the conditions for a message to be sent and received through a copper cable. His highly renowned {\it Information Theory} made it possible to study real systems where the concept of energy was not applicable, or important, or central. Because, after all, the money or the goods that are exchanged in the economic processes of buying and selling, due to their intrinsic characteristics, have become {\it equivalent} to energy, but {\it are not energy}. The advantage of working with entropy is that this is a property that can be defined and calculated without the need to define energy.

Another issue that has been widely studied in the literature is the origin and evolution of {\it prices}, as an essential element in the formal description of an economic system where merchandise is bought and sold. The majority economic thought tells us that when prices increase then the quantity demanded decreases, and vice versa (``Law of demand"\ ). This behavior between the quantity of goods and prices, from a strictly conceptual point of view, as far as we know, has not been sufficiently addressed by science. Part of the motivation of this work is translated into the need to have a clear and simple method to relate these two quantities, within a coherent theoretical framework. In this way we could have a way, at least qualitatively, of ``redescovering"\ the properties that should be observed in real economic systems when the goods to be exchanged is money itself. In other words, when the market for buying and selling is the currency of a country, which is purchased with the currency of another country, there is interest in learning about the mechanisms that appear when money is printed and released into the torrent of the economy. And this is the second motivation of this work.

Finally, and as occurs in physical thermodynamic systems, knowing the equivalence between temperature, or its inverse, with some factor in the statistical description of the probability of occupation of the levels or states of the agents, is an essential step to make the macroscopic and statistical descriptions of these systems are also equivalent.

\section{Economic Model}\label{probabilidad-ocupacion-estados-k}

The way we will proceed is equivalent to when we are study in statistical mechanics the probability of finding
a system, within a set of systems, with energy $E_k$. Here, however, we will not move in space of the energies
$\{E_1, E_2, E_3, ...E_k,...E_m\}$, but in the "space of goods" $\{\mathbf{Q}_1, \mathbf{Q}_1, \mathbf{Q}_3, ...,\mathbf{Q}_k,...\mathbf{Q}_m\}$, where each of the N economic agents, which simply we will call {\it agents}, possesses a certain quantity
of goods $\mathbf{Q}_1$, $\mathbf{Q}_2$, $\mathbf{Q}_3,...,\mathbf{Q}_k,...\mathbf{Q}_m$, product of exchange between them, which we will consider here simulate the buying-selling process in a real system. The amounts $\mathbf{Q}_k$ 
belong to the natural numbers, including zero, and have a maximum value $\mathbf{Q}_m$ 
\footnote{This is the so-called Upper Limit or "Roof".}. 
That is, the space like this
created is finite. And also it is fulfilled $\mathbf{Q}_1$ < $\mathbf{Q}_2$ < $\mathbf{Q}_3 <...<\mathbf{Q}_k <...<\mathbf{Q}_m$.

We will understand by "goods" o "merchandise" those ones that can be measured, numbered, accumulated, divided, added and exchanged by human being, like fruits, vegetables, cereals, gold, oil, currencies (physical or electronic), cryptocurrencies, stocks corporate, etc., and therefore can be associated with a {\it price}, within a scale, as a reflection of its {\it value}.

We will assume that the buying and selling of merchandise is completely random, and that, as is logical to consider, in this process there is also an exchange of money. However, we are only interested in the distribution of merchandise, since in this dynamic the {\it price} at which they are traded {\it emerges naturally}. Agents do not seek any benefit, nor minimize and/or maximize any amount, nor do they follow cost-benefit strategies. It is also reasonable to consider that here those agents who, at any particular moment, do not own merchandise should not be excluded from the dynamics, since they can continue to participate in the economic dynamics by obtaining them later by appropriation, donation, deceit, persuasion, work, etc

Within the so-called {\it commodities}, which are basic merchandise, such as agricultural products and minerals (gold, iron, etc.), it is possible, formally, to consider a special type of commodity: money \cite{Yakovenko2013-2}, despite being a product created by human beings, and in that sense be different from other consumable goods such as food, fuel, etc. That is, the money itself (we are considering it of the {\it fiat} type, even when money backed by gold, silver, etc., are not excluded), when being exchanged between the agents, also generates a purchase-sale price, which may be expressed in another currency that enjoys greater stability and credibility, or on commodities such as gold, diamonds, etc.

Let us now characterize the economic system thus constructed:
the total number of goods Q in this economic system is simply
\begin{equation}
Q = \sum_{k=1}^m\, n_k\,\mathbf{\overline {Q}}_k\,,
\label{10}
\end{equation}
where
\begin{equation}
\mathbf{\overline {Q}}_k = (\mathbf{Q}_k/n_k)\,.
\label{20}
\end{equation}
$n_k$ is the number of agents that owns $\mathbf{Q}_k$ goods each, and $1 \leq m \leq N$. $n_k$ are also known as {\it Occupation numbers}.

It must be true, of course, that the total number of agents in the system is
\begin{equation}
N = \sum_{k=1}^m n_k\,.
\label{30}
\end{equation}

The total quantity of goods $Q$ and the total number of agents $N$ are fixed quantities. In the system thus constructed, 
there is no creation or destruction of agents or goods.

As an example, let us consider that the goods are bananas. In this society of 10 people who exchange, buying or selling, bananas with each other, the total number of bananas is, for example, 25. At some given moment, 4 people each have 1 banana; 3 people each have 2 bananas; 2 people each have 4 bananas; and 1 person has 7 bananas. Then it is fulfilled:
$Q = \sum_{k=1}^m\,n_k\,\mathbf{\overline {Q}}_k = 4(4/4) + 3(6/3) + 2(8/2) + 1(7/1) =
4(1) + 3(2) + 2(4) + 1(7) = 25$.

$N = \sum_{k=1}^m\,n_k = 4 + 3 + 2 + 1 = 10$.

Note that the total number of agent groups or levels is $m = 4$.

\section{Probability calculation}

According to combinatorial theory \cite{McQuarrie1}, the number of ways $w$ to distribute $N$ agents among levels 
 $n_1, n_2, n_3,...,n_m$ is
\begin{equation}
w = \frac{N!}{\prod_{k=1}^{m}\,n_{k}!}\,,
\label{40}
\end{equation}
where $\Pi_{k=1}^{m} n_k! = n_1\,!n_2!\,n_3!\,...\,n_k!\,... $ is the productoria of the  $n_k!$. The symbol $!$ represents the factorial of the respective term.

Let's maximize \footnote{Since $w$ is normally a fabulously large number, it is wise to maximize $\ln w$ instead of $w$. It is possible to show that the maximum of $\ln w$ coincides with the maximum of $w$.} 
$\ln w$ with respect to the $n_k$, subject to Eqs. \ref{10} and \ref{30}. To do this, we will use the Lagrange Indeterminate Multipliers Method (MIL) \cite{McQuarrie2}, which allows maximizing a function subject to conditions, thus obtaining
\begin{equation}
\frac{\partial}{\partial n_k}\Bigl(\ln\,w-\alpha \sum_{i=1}^m n_i -\beta \sum_{i=1}^m n_i\,\overline {Q}_i\Bigr) = 0; k = 1,2,3,..,m \,.
\label{50}
\end{equation}
Since there are two conditions or restrictions imposed on the function $\ln\,w$, then there are two constants, $\alpha$ and $\beta$.

Using the above equation, it is found that
\begin{equation}
-\ln\, n_k -1 -\alpha-\beta \overline {Q}_k = 0\,,
\label{60}
\end{equation}
where we have used the Stirling approximation \cite{McQuarrie3} $\ln x!\approx x\ln x - x$. $\ln$ is the natural logarithm to base $e$.

If the probability of finding $n_k$ agents with number of goods $\overline{Q}_k$ is
\begin{equation}
p_k\, =\, n_k/N\,, 
\label{70}
\end{equation}
then from Eq. \ref{60} we can write
\begin{equation}
p_k = C\,e^{-\beta \overline {Q}_k}\,,
\label{80}
\end{equation}
where 
\begin{equation}
C^{\,-1} = \sum_{l=1}^m\, e^{-\beta \overline {Q}_l}\,.
\label{90}
\end{equation}

For the reader familiar with statistical mechanics, $p_k$ is the probability that, taking a random system from the set of physical systems, it is in the energy state $k$, with energy $\overline{E} _k$. Similarly, $C$ would be the inverse of {\it partition function}; which ensures that the probabilities are {\it normalized}, $\sum_k p_k \equiv 1$.

If we focus on any of the systems within the set of physical systems, it is also possible to show that the probability $p_k$ of finding a particle with energy $\epsilon_k$ in a set of $N$ particles at temperature $T$ is $ p_k = C\,e^{-\beta {\epsilon}_k}$, where the inverse of the normalization constant is $C^{\,-1} = \sum_{l=1}^m\ , e ^{-\beta{\epsilon}_l}$. This is the well-known Boltzmann probability distribution \cite{Chandler5}.

$\beta$ is a fundamental quantity that can only be determined unambiguously by comparing it to the physical world. In thermodynamic systems, it can be shown that $\beta \equiv 1/k_BT$ \cite{Chandler}, where $T$ is the absolute temperature measured in kelvins, and $k_B$ is the Boltzmann's constant. Identifying $\beta$ as the inverse of temperature allows us to pass, fluidly, from the statistical description to the classical description, and vice versa.
In statistical mechanics, using first principles, there is no way to correctly determine the identity of $\beta$, without appealing to classical thermodynamics. Starting from this relationship, it is transparent to move coherently between both descriptions.

In this economic system $\beta$ must also be determined, as we will do later.
\section{Total revenue in the purchase-sale process}\label{derivacion precio}

Let us suppose for simplicity, but without losing generality, that this system of economic agents buys and sells a single type of good. Now consider plausibly that the {\it Total Revenue} $P$ \cite{Landsburg}
to be received by the seller, which is the same amount of money that will be paid to the buyer,
is directly proportional to the total quantity of goods sold
$Q$, $P \propto Q$.

Going from a proportionality to an equality, it will be necessary to multiply by a constant, which we will call $b$. We will call this amount {\it unit price} of the goods.

As the quantity of goods $Q$ cannot be a negative number, then the sign of $P$ is the same sign of $b$, which
{\it is usually} positive
\footnote{In April 2020, an unprecedented event occurred, when the price of WTI oil registered a negative value. That specific case, which occurred in the context of global combat
against the Covid-19 virus, demonstrating that, in certain circumstances, the price of a merchandise
could become negative.}.
Therefore
\begin{equation}
P = b\,Q\,.
\label{100}
\end{equation}

For example, if $10$ bananas are bought and sold, at a price of $0.25$ {\it currency units um} each, then the buyer will deliver to the seller $P = b\times Q = 0.25$ $um/unit \times 10\,units = 2.5\,um$. If the price $b$ is fixed, that is, it is independent of the quantity of goods in the purchase and sale, then it is said that we are in a {\it Perfectly Competitive} (PC) market \cite{Landsburg3}. If, on the other hand, the price depends on the amount of merchandise traded, then we are talking about a {\it Monopolistic Competition} (MC) market  \cite{Landsburg4} .

\begin{figure}
	\centering
 \includegraphics[width=12cm]
    {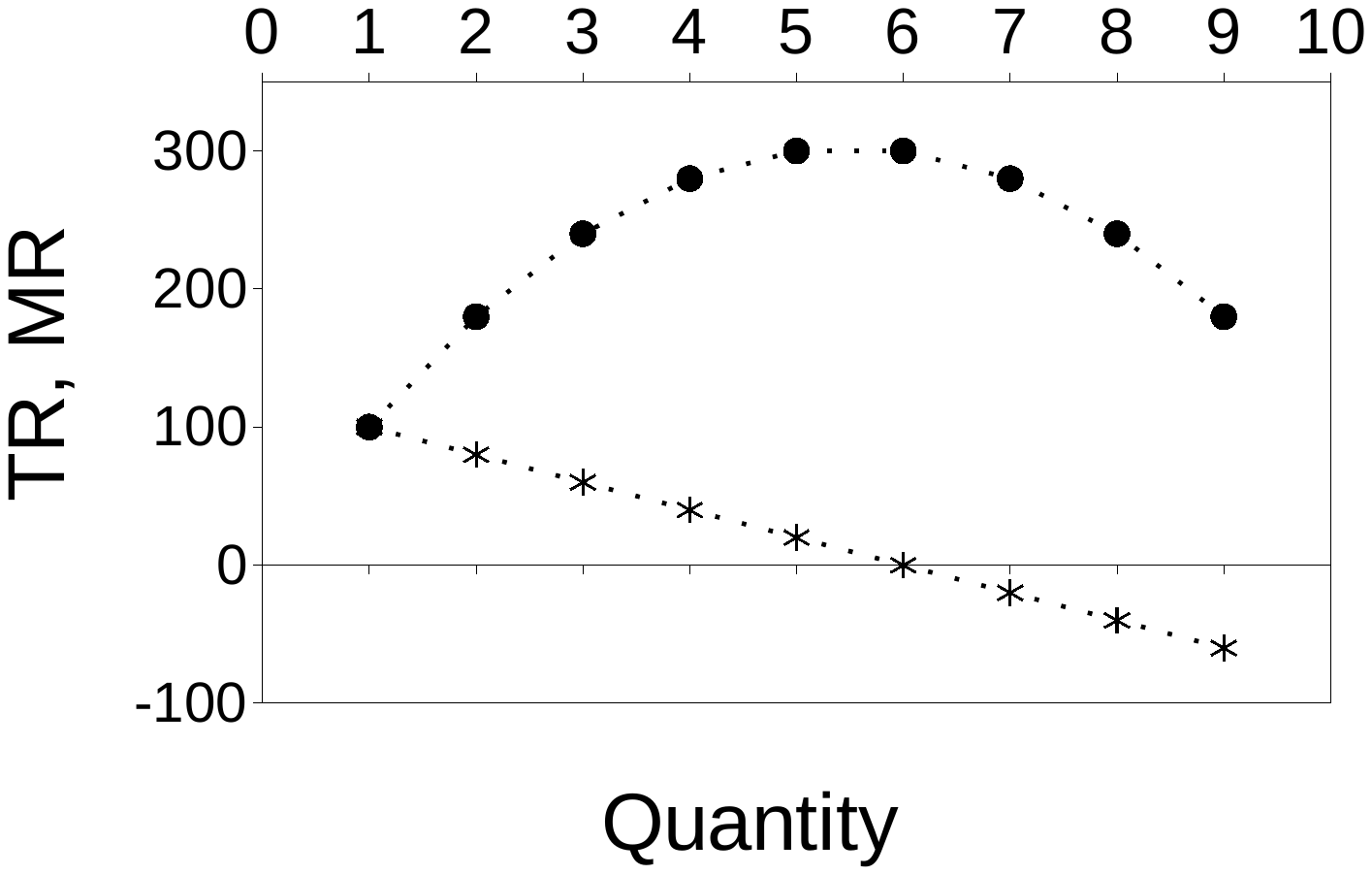}
 \caption{Total Revenue ($\bullet$) and Marginal Revenue ($\ast$) vs Quantity sold, according 
 to the data in Table \ref{Total-Revenue}.}
 \label{TRMRLIBRO-VS-Q}
\end{figure}

The table \ref{Total-Revenue} shows \footnote{This table is equivalente to the Landsburg teaching example \cite{Landsburg5}.}, by way of example, the calculation of Total Revenue $P$ and {\it Marginal Revenue} $P_M$, understood as the change in $P$ due to the effect of the change in the amount of goods $Q$, $P_M = \Delta P / \Delta Q$, when a product was bought-sold at certain prices. The general issue is that, in order to be able to sell larger volumes of goods, the seller gradually reduces the unit price, so that, for example, when the price is $80$, and therefore $3$ are sold units of the product, then it is obtained from the sale $P = $ price $\times$ Quantity of Goods $= 80 \times 3 = 240$, calculated in a certain monetary unit $um$. In the latter case, $P_M = (240 - 180) / (3 - 2) = +\,60$. But, if the price is lowered to $30$ in order to sell more merchandise, in this case $8$, then $P = 30 \times 8 = 240$, and the Marginal Revenue $P_M = (240 - 280) / ( 8 - 7) = -\,40$. Clearly in the first case, from the point of view of the seller, it is a convenient situation, since it obtains higher income ($P_M > 0$). In the second case, always from the point of view of the seller, he is losing money, since the total revenue $P$ has decreased ($P_M < 0$), despite selling more merchandise. It is obvious that the maximum profit is achieved at the maximum of $P$, where $P_M = 0$. Fig. \ref{TRMRLIBRO-VS-Q} shows the Total Revenue and the Margial Revenue, as a function of the quantity of goods sold, according to the table \ref{Total-Revenue}. Note that $P_M$ changed sign at the maximum of $P$.

\begin{table}
 \centering
 \caption{Examples of Total Revenue and Marginal Revenue.}

 \begin{tabular}{c|c|c|c}
  $Quantity$&$Price$&$Total\, Revenue$&$Marginal\, Revenue$\\
  \hline
  1&100&100&+100\\
  2&90&180&+80\\
  3&80&240&+60\\
  4&70&280&+40\\
  5&60&300&+20\\
  6&50&300&0\\
  7&40&280&-20\\
  8&30&240&-40\\
  9&20&180&-60\\
 \label{Total-Revenue}
      \end{tabular}

\end{table}

Since we are interested in working with quantities that are independent of a particular monetary unit $um$, the scale on which we will be measuring $P$, in its various forms, will be on a scale relative to a {\it monetary unit of reference $um_0$}, which we will consider fixed and stable over time. For the case $um_0$ we can write $P_0 = \mathsf b_0\,Q_0$, where $P_0$ is the Total Revenue in $um_0$ of $Q_0$ goods, $b_0$ is the unit price in $um_0 $, and $Q_0$ is the quantity of goods, which we will take as the unit. Since the quantity of goods is a number belonging to the natural numbers, including zero, in any of the scales - $um$ or $um_0$ - it must have the same meaning: three bananas are three bananas on both scales, ten red balls are ten red balls on both scales, etc.

Let us then take the quotient of the expressions for Total Revenue in $um$ and $um_0$, $(P / P_0) = (b / b_0) \,(Q / Q_0)$. From now on, to simplify the nomenclature, $(P / P_0)$ will be called $P$, $(b / b_0)$ will be called $b$, $(Q / Q_0)$ will be called $Q$ . It is then clear that $P$, $b$ and $Q$ are now {\it dimensionless}, and would be measured in $um$ relative to their value in the reference $um_0$. Therefore Eq. \ref{100} is still $P = b\,Q$. Viewed this way, $b$ is a {\it relative price} \cite{Landsburg2}.

If we derive Eq. \ref{100} with respect to the number of agents $n_k$, keeping the other $n_j$ different from $n_k$ constant, it is fulfilled
\begin{equation}
\overline P_k = b\,\overline Q_k\,,
\label{160}
\end{equation}
where 
\begin{equation}
\overline P_k = \Bigl(\partial P/\partial n_k\Bigr)_{n_{j\neq k}}\,,  
\label{170}
\end{equation}
\begin{equation}
\overline {Q}_k = \Bigl(\partial Q/\partial n_k \Bigr)_{n_{j\neq k}}\,.
\label{180}
\end{equation}
$\overline{P}_k$ is the Total Revenue when associated with the sale of $\overline{Q}_k$ amount of goods.

If we now take Eq. \ref{160}, and calculate the statistical average over all possible values of $k$, we obtain $\overline P = b\,\overline Q$ for the average values of $\overline P = P /N$ and $\overline Q = Q / N$, where
\begin{equation}
\overline P = 
\sum_{k=1}^m\,p_k\,\overline P_k 
\,,
\label{190}
\end{equation}
\begin{equation}
\overline Q =
\sum_{k=1}^m\,p_k\,\overline Q_k 
\,.
\label{200}
\end{equation}
$p_k = n_k / N$ is the probability, given by Eq. \ref{80}, $\overline{P}$ and $\overline{Q}$, is the {\it Total Revenue} and the {\it Total Quantity of Goods per agent}, respectively.

Strictly speaking, if we want to do a statistical average over all values of $k$, Eq. \ref{160} would have to be rewritten as $\overline P_k = (b\,\overline Q)_k$, which means that the product $(b\overline{Q})_k$ would have a given value for each $k$. If we write $(b\,\overline Q)_k \approx b \overline{Q}_k$, we are considering that $b$ is a constant and that it is the same for any value $\overline{Q}_k$, be it is it big or small. In other words, the unit price $b$ is the same for any commodity value, whether it is {\it retail} ($\overline{Q}_k$ small) or {\it wholesale} ($ \overline{Q}_k$ large).

\section{Euler theorem for homogeneous functions}\label{teorema euler}
Let's compare the last two expressions with the {\it Euler's Theorem} for homogeneous functions \cite{McQuarrie4, Chandler2}, better known in Physical chemistry as {\it Additivity Rule} \cite{Castellan1} in the determination of the {\it Partial Molar Quantities} \cite{Lewis}
\begin{equation}
X = \sum_{k=1}^{m} n_k\,\overline X_k\,;
\label{210}
\end{equation}
where an extensive property $X$, Euler function of degree $1$, can be expressed in terms of the corresponding intensive properties $\overline X_k$ \footnote{A characteristic of a {\it intensive property} is that it is constant throughout the entire system, and is obtained by dividing two extensive properties.}, Euler function of degree $0$
\begin{equation}
\overline X_k = \Bigl(\partial X / \partial n_k\Bigr)_{n_{j\neq k}}\,.
\label{220}
\end{equation}

It should be remembered that extensive properties \cite{Chandler3} are additive, so that the total value in a given system is the sum of this property in each of the $m$ parts into which it is divided. For physical systems, volume, amount of matter, energy, entropy, etc., are examples of extensive properties. Intensive properties, on the other hand, cannot be added, as is the case of temperature, chemical potential, density, etc., and their value is equal and constant in any of the $m$ parts of the system. It must also be taken into account that the quantities $\overline X_k$ {\it can be positive or negative}. For example, in physicochemical systems, the molar volume of a pure substance (that is, the volume occupied by one mole of substance) is always positive, but the partial molar volume of that same substance in a mixture with other substances, it can be positive or negative. The same is true for the partial molar heat capacity \cite{Lewis2}.

If the values of $\overline{X}_k$ {\it are independent of each other}, as would happen, for example, with the molar volume when mixing two gases that behave in an ideal way, then we can ensure that \footnote{In a mixture the {\it partial molar volume} $\overline{V}_k = \Bigl(\partial V / \partial n_k\Bigr)_{n_{j\neq k}}$ of component $k $, which is the Euler property of degree $0$, is equal to its molar volume $V_k / n_k$, only when the volume of each component in the mixture does not depend on the volume of the other components. Here $V_k$ is the volume occupied by $n_k$ moles of component $k$.}
\begin{equation}
\Bigl(\partial X /\partial n_k\Bigr)_{n_{j\neq k}}
\cong
\Bigl(X_k / n_k \Bigr)\,.
\label{230}
\end{equation}
Substituting the previous expression in Eq. \ref{210}, it is shown that, as expected, the extensive property $X$ is equal to the sum of the property in each of the $m$ parts in which the system is composed, $X =\,\sum_{k=1}^{m}\,X_k$. What is clear is that, in any case, what exactly satisfies the Additivity Rule, Eq. \ref{210}, are the Euler functions of degree $0$, $\overline{X}_k$. In the case of the economic system studied here, $P$ and $Q$ are Euler functions of degree $1$, and $\overline{P}_k$ and $\overline{Q}_k$ are Euler functions of degree $0 $.

From the probability $p_k$, Eq. \ref{80}, we can write $\beta\,\overline {Q}_k = -\,\ln p_k + \ln C$. Taking now the statistical average of $\beta\,\overline Q_k$, we have
\begin{equation}
\overline P = 
\,\sum_{k=1}^m\,p_k\, \beta\,\overline Q_k
\,,
\label{280}
\end{equation}

\begin{equation}
\overline{P}
=
-\,\sum_{k=1}^m\,p_k\ln p_k + \,\ln C
\,.
\label{300}
\end{equation}
Where we have taken $\sum_{k=1}^m\, p_k = 1$.

For Eq. \ref{280} to be consistent with Eqs. \ref{160} and \ref{190}, the constant $\beta$ will be identified with the unit price $b$
\begin{equation}
\beta \equiv b
\,.
\label{290}
\end{equation}

The latter is in line with what was done by \cite{Foley1994}, when he identified ${\bf \Pi}$ as the {\it vector of entropic prices}, when calculating the probability $p( {\bf x}_k) = N_k/N = C e^{- {\bf \Pi}\,.\,{\bf x}_k}$ to find an agent $k$ in an economic system, which is in a state ${\bf x}_k$, where $C$ is the normalization constant. The vector ${\bf x}_k$ can be considered as an n-dimensional generalization of the energies $\epsilon_k$ when studying the probability distribution $p$ in a system of $N$ molecules, $p(\epsilon_k) \propto e^{- \beta\,\epsilon_k}$, where the Lagrange multiplier $\beta$ is inversely proportional to the thermodynamic temperature $T$.

\section{Configurational Entropy and Information}\label{entropia configuracional}
In physical systems, the {\it average configurational entropy} in reduced units \footnote{This is strictly true only when $N\,\gg\,1$, and for systems of $N$ independent and distinguishable particles. 
\cite{Zupanovic} made an interesting proof, concluding that the most probable value of Boltzmann entropy in reduced units (${S} ^{\,*} = \ln\,w$), is $N$ times the Gibbs entropy in reduced units ($\overline {S}^{\,*} = - \sum_k\,p_k \,\ln p_k$), in the limit $N\,\gg\,1$. $\overline {S}^{\,*} = S^{\,*}/N$, where ${S}^{\,*}= S/k_B$.}, for $N$ agents who distribute a total energy $E,$ distributed in $m$ groups or levels, it can be written as
\begin{equation}
\overline {S}^{\,*} 
=\,-\,\sum_{k=1}^m\,p_k\,\ln p_k\,.
\label{310}
\end{equation}

This expression is equivalent, for sufficiently large $N$, to the well-known Boltzmann equation $\overline {S}_k^{\,*} = \ln w / N$, where $w$ is calculated with Eq. \ref{40}. In fact, by applying MIL to $\ln w$, what is found is the probability $p_k$ that maximizes the entropy of the system \cite{Bean5}. In physical systems, this is equivalent to minimizing the energy \cite{Shavit}.

Entropy in physical systems is an extensive, state, {\it emergent}, non-mechanical, and statistical property. But the concept of entropy can be applied to systems of {\it many bodies} where the concept of energy does not exist, is not defined, or does not need to be used. In economic and social systems, in literary studies, etc., entropy can be applied and calculated, as a measure of the {\it Quantity of Information} that can be extracted from the system.

In 1948 Shannon extended the {\it Second Law of Thermodynamics}, that is, entropy, to non-physical systems, through his {\it Information Theory} \cite{Shannon2}, also known as {\it Communication Theory}. It derives the function call {\it Information} $I$ \footnote{The logarithm could have any base, mostly $10$, $e$ or $2$.}
\begin{equation}
I 
= \,-\,K\,\sum_{k=1}^m\,p_k\,\log p_k\,,
\label{315}
\end{equation}
where $K$ is a positive constant, which measures the information, the {\it choice} and the {\it uncertainty} of the system. It is clear that $\overline {S}^{\,*} = I$, where, without loss of generality, we have taken $K = 1$ \cite{Bean3} and logarithm base is $e$.

In this context, entropy, or equivalently Information, can be considered as a measure of the uncertainty in the description of the economic system presented here.

\section{Unit price}\label{ingreso-total-entropia}

Comparing the Eqs. \ref{300} and \ref{310}, it is clear that we can write
\begin{equation}
\overline P\,
= \,\overline {S}^{\,*} + \ln C (\overline{Q})\,.
\label{320}
\end{equation}
It is easy to prove, remembering that $p_k= n_k/N$, which also hold \footnote{The term $C(\overline{Q})$ implies that the constant $C$ depends on the value of $\overline Q$. When $C$ is written it means that it takes a single value for all $\overline{Q}_k$.}
\begin{equation}
\overline P_k\,
= \,\overline {S}_k^{\,*} + \ln C\,,
\label{330}
\end{equation}
\begin{equation}
\overline {S}^{\,*} 
=\,\sum_{k=1}^m\,p_k\,\overline{S}^{\,*}_k\,,
\label{340}
\end{equation}
\begin{equation}
\overline {S}^{\,*}_k 
=\,-\,\ln p_k\,.
\label{350}
\end{equation}

In the different interpretations of the Information given in the literature \cite{Bean}, the term $-\,\log p_k$ \footnote{For the analysis it is irrelevant to use any base for the logarithm.} is a measure of ``unlikely"\ , ``surprise"\ or the ``unexpectedness"\ , then $\overline {S}^{\,*} = -\,\sum_k p_k\,\ln p_k$ a measure of the {\it average} of improbability, surprise, or unexpectedness, given number of events or measurements. Also, the term $-\ln p_k$ is sometimes called `` self-information"\ or ``information content"\ \cite{Applebaum2}.

From the Eqs. \ref{320} and \ref{330} then we have two equivalent expressions for the unit price $b$
\begin{equation}
b\,
= \,(\overline {S}^{\,*} + \ln C(\overline{Q}))/\overline{Q}\,,
\label{355}
\end{equation}
\begin{equation}
b\,
= \,(\overline {S}_k^{\,*} + \ln C)/\overline{Q}_k\,.
\label{360}
\end{equation}
From the above equations, it is true that \footnote{This is so because $\overline{Q} = \Bigl(\partial \overline {A}^{\,*} / \partial b\Bigr)$.}
\begin{equation}
b\,
= \Bigl(\partial \overline {S}^{\,*} / \partial \overline{Q}\Bigr)
\,,
\label{365}
\end{equation}
\begin{equation}
b\,
= \Bigl(\partial \overline {S}_k^{\,*} / \partial \overline{Q}_k\Bigr)_{{j\neq k}}\,.
\label{370}
\end{equation}

\section{Thermodynamic Euler functions}\label{euler-termo}
Let us compare Eq. \ref{320} with the equation of thermodynamic origin, for a system of $N$ particles at temperature $T$ and volume $V$
\begin{equation}
\overline {E}^{\,*}\,=
\overline {S}^{\,*} + \overline {A}^{\,*};
\label{361}
\end{equation}
where $\overline {E}^{\,*} = E^{\,*}/N$ is the {\it relative} energy per particle, $\overline {A}^{\,*} = A^{\,*}/N$ 
is the {\it relative} Helmnholtz free energy per particle
\footnote{Where $E^{\,*}= \beta E$ and $A^{\,*}= \beta A$.}, 
$\overline {S}^{\,*} = {S}^{\,*}/N = - \sum_{k=1}^m\ ,p_k\,\ln p_k$ is the reduced entropy per particle (where ${S}^{\,*} = S / k_B$), $p_k = Z^{-1} e^ {-\beta \overline {E}_k}$ is the probability of finding a system with energy $\overline{E}_k$, $Z = \sum_{k=1}^m\, e^{ -\beta \overline {E}_k}$ is the partition function, and $\beta\, = 1/k_B\,T$. From statistical mechanics it is also known that \cite{Bromberg} $\overline {A}^{\,*} = - \ln Z$.

Since $E^{\,*}$, $S^{\,*}$ and $A^{\,*}$ are Euler functions of degree 1 and therefore satisfy expressions of the type $X = \sum_{k=1}^m\,n_k\,\overline X_k $, from the previous expression we can ensure that
\begin{equation}
\overline {E}^{\,*}_k\,=
\overline {S}^{\,*}_k + \overline {A}^{\,*}_k\,.
\label{372}
\end{equation}
Therefore, the {\it Average Total Revenue} per economic agent $\overline{P}$ is {equivalent}, in physical systems, to the energy $\overline {E}^{\,*}$, and has two contributions: a configurational entropy $\overline {S}^{\,*}$ and another of the free energy type $\overline {A}^{\,*} = \ln C(\overline{ Q})$.

Note that by Euler's Theorem \ref{210}, also $\overline{P}_k$ is equivalent to $\overline {E}^{\,*}_k$, and $\ln C$ is equivalent to $\overline {A}^{\,*}_k$.

\section{Absolute negative temperatures}\label{temperaturas negativas}

When studying thermodynamic systems, it is assumed that $\beta=1/k_BT$ is always positive. However, this is not strictly correct for systems where the energy states accessible to the system have a "roof" of energy $E_m$.

\begin{figure}
	\centering
 \includegraphics[width=12cm]
{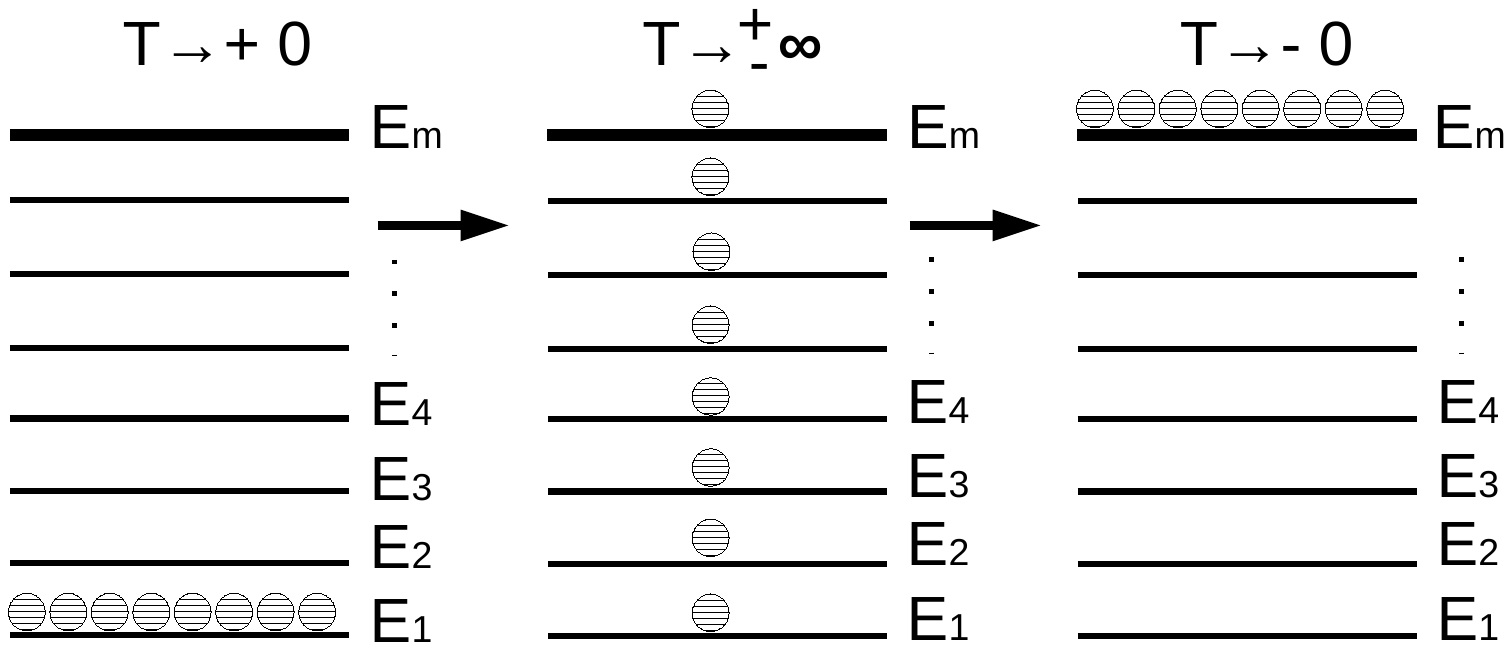}
\caption{Representation of a system where the number of energy levels is limited by a ``roof"\ $E_m$.\label{temperaturas-negativas-sistemas-techo}}
\end{figure}

Let's conceive a system where energy levels are limited. The vast majority of classical systems are not bounded or limited in terms of the maximum energies that they can acquire. However, there are quantum systems which show this bounded energy. In this case, and as illustrated schematically in Fig. \ref{temperaturas-negativas-sistemas-techo}, as the energy in the system increases, the states are populated of greater energy until reaching a situation that we can define as ``infinite temperature", in which it is true that the occupancy numbers of the level $k$ tend to the unit $n_k \approx 1$. However, if we continue to increase the energy, and as the accessible levels are upper-bounded, the particles gradually "accumulate"\ in the level of highest energy $E_m$.

At $N$ and volume $V$ constant, there is an exact relationship between entropy $\overline {S}^{\,*}$, energy per particle $\overline {E} = E / N$, and temperature $T$ \cite{Chandler4}
\begin{equation}
(k_B\,T)^{-1} = \Bigl(\partial \overline {S}^{\,*} / \partial \overline E \Bigr)\,.	
\label{420}
\end{equation}

For the classical case -without ``roof"- starting from zero entropy, it is clear that the entropy will grow monotonically with the energy, so that then the temperature, and therefore $\beta$, will always be positive. For the quantum case -with ``roof"- the temperature will always be positive, before reaching ``infinite temperature", since the entropy is increases as the energy increases, as occurs in the classical case, at maximum $T\to \pm \infty$ and $\beta \to \pm 0$.
But then, as the particles ``accumulate"\ at the highest energy level, the entropy clearly decreases as the energy continues to increase, until to reach zero entropy again, then the temperature $T$, and therefore $\beta$, would now be negative.

\begin{figure}
	\centering
 \includegraphics[width=12cm]
    {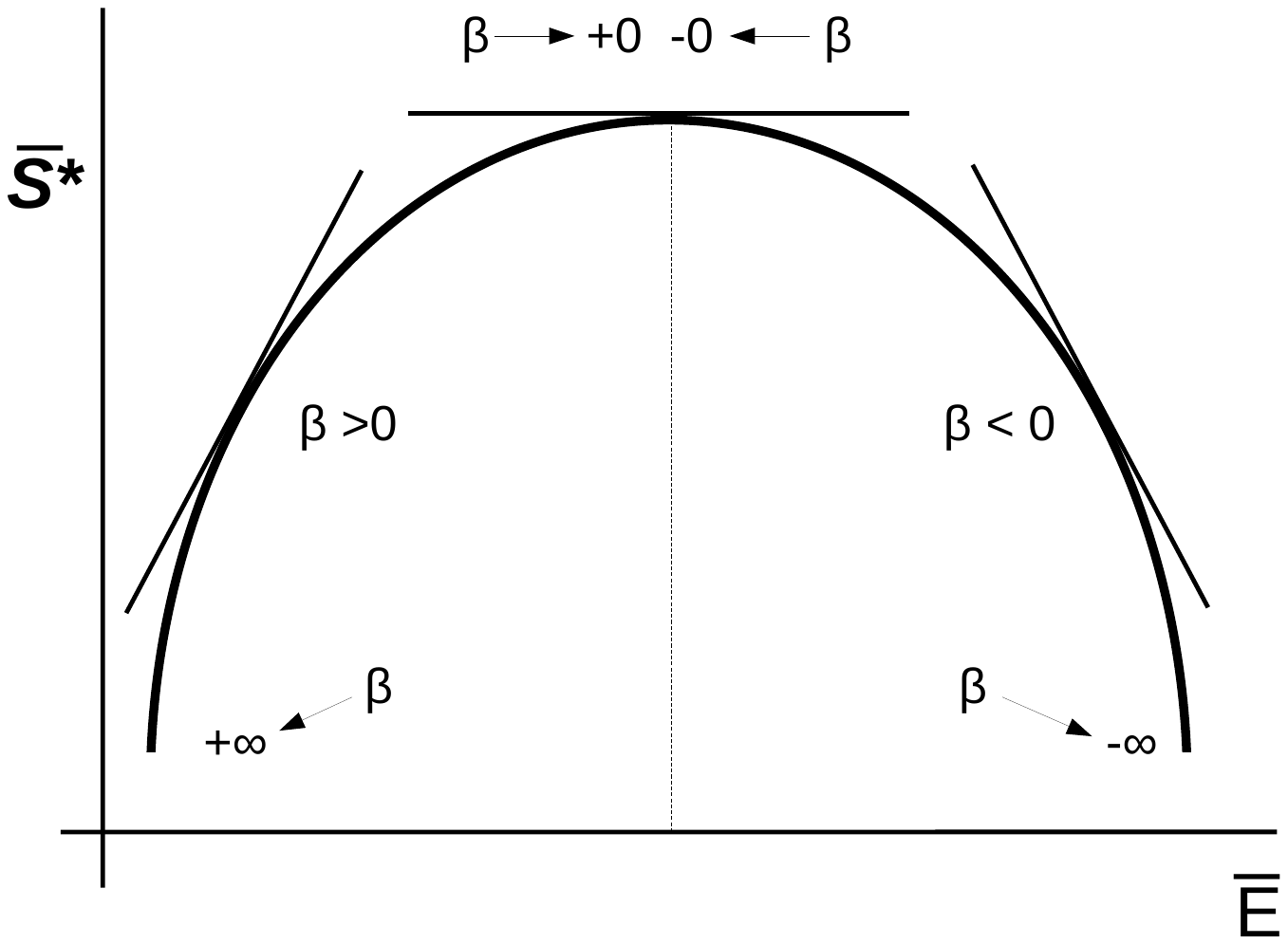}
 \caption{Schematic representation of the entropy $\overline {S}^{\,*}$ vs the energy $\overline E$ of a thermodynamic system, where the number of energy levels is finite.
 \label{SvsEnegativa}
}
\end{figure}

The existence of negative absolute temperatures has a solid theoretical basis \cite{Landau}, being observed experimentally in quantum systems with a limited number of energy states \cite{Purcell} and conceptually clarified by 
\cite{Ramsey} in the 50s of the last century. These types of temperatures have also been observed in experiments as dissimilar as the ultracold bosons \cite{Braun} and in the Axelrod model of cultural convergence \cite{Villegas-Olivares}.

Figure (\ref{SvsEnegativa}) schematically represents the entropy $\overline {S}^{\,*}$ as a function of energy $\overline E$, for a system with a finite number of states. The slope, point to point, is $\beta = 1 / k_B\,T$.

One of the consequences of this type of temperature is that the {\it Absolute temperature scale} must also contain negative values:
\begin{equation}
+0\,K...+300\,K...+\infty\,K...-\infty\,K...-300\,K...-0\,K\,.
\label{430}
\end{equation}
Note that the lowest possible temperature is still $0\,K$, but it is preceded by the symbol ``+", that is, the ``positive zero", $+0$, to differentiate it from $-0$, or ``negative zero". This $+0$, would be the zero that we are ``accustomed"\ to working with. Starting from $0\,K$, as we increase the energy of this system, its temperature reaches higher and higher values (always positive), until we reach the ``plus infinity", $+\infty\,K$. This $+\infty\,K$ would be the infinity we are "used" to working with. From a statistical point of view, each of the particles in the system now occupies a different energy state. If we now continue to increase the energy of the system, the particles will necessarily "accumulate" ("condense") at the maximum energy level (the ``roof"\ energy). From that moment the temperatures will all begin to be negative, and therefore the {\it inversion of the population} will begin to be observed and it will begin to drop from $-\, \infty\,K$
\footnote{The temperature $-\,\infty\,K$ is identical to $+\,\infty\,K$ and both give the same distribution and the same values of thermodynamic quantities of the system.}
until reaching a hypothetical situation where all the particles would be at that maximum energy level. There the temperature will return to be zero kelvin, but negative, $-0\,K$.

That is, negative temperatures are much ``warmer" than positive ``normal"\ temperatures. In other words, the ``true zero"\ , that is, the lowest possible temperature, is still $+\,0\,K$. The negative zero, $-\,0\,K$, then, turns out to be much ``hotter"\ than infinity, positive or negative.

Finally, as demonstrated by 
\cite{Dunn}:
a) Equilibrium states with negative temperatures {\it are possible}.
b) Those states with infinite temperature have the greatest total entropy.
c) Those states with negative temperatures have a greater total energy than any state with positive temperature.

\section{Price determination in $k$}\label{El precio}
Let's look at $\overline {P}_k$, the total revenue per agent, when $\overline{Q}_k$ quantities of merchandise are traded. For example: in a given $k$, where there are two agents, $n_k = 2$, a total of 10 USD was paid for 20 pens. Here, the Total Revenue on the site $k$, $P_k = 10$. The total quantity of merchandise $Q_k = 20$. These two amounts per economic agent are: $\overline{P}_k = 10 / 2 = $5 and $\overline{Q}_k = 20 / 2 = $10. If we divide 10 by 20 pens, 10 / 20 = 0.5, we get the price per merchandise; that is, the unit price $b$. So if we divide $\overline{P}_k$ by $\overline{Q}_k $, we get, as expected, the unit price $b = \overline {P}_k / \overline{Q}_k$ .

If we now ask ourselves, {\it statistically}, about the probability of finding, in the entire economic system, a given Total Revenue  per agent $\overline {P}_k$ (and/or merchandise $ \overline{Q}_k$), we get the {\it Total Expected Revenue} $\overline {P\,'}_k = p_k\, \overline {P}_k$.

If we now divide $\overline {P\,'}_k$ by the quantity of goods $\overline{Q}_k$, we finally obtain the {\it Price} $\overline{P}^{\,* }_k$
\begin{equation}
\overline{P}^{\,*}_k 
=b\, C e^{- b \overline{Q}_k}
\,.
\label{390}
\end{equation}

That is, the price $\overline{P}^{\,*}_k$, in reduced units {\it um/umo}, is nothing more than the unit price determined statistically considering {\it the entire} economic system. Note that if $\overline{Q}_k \to 0$, then the price $\overline{P}^{\,*}_k \to b\,C$. In other words, when we study a single value of $k$, the unit price is $b$. But, when all the values of $k$ are considered statistically, then the unit price {\it effective} is the product $b\,C$. Observe that the greater the amount of merchandise $\overline{Q}_k$ is traded, the lower the price given the exponentially decreasing term in Eq. \ref{390}. In other words, in a {\it natural way}, high prices are predicted when very few merchandises are traded (what is known as the "retail price"), and lower prices when more merchandise is traded or exchanged (''wholesale price``).

In short, what Eq. \ref{390} tells us is that the price $\overline{P}^{\,*}_k$ of a commodity or goods is {\it exponentially decreasing} with the quantity of goods involved, as long as $b > 0$ \footnote{In very rare cases, of academic value only, where the unit price $b$ is a negative number, $\overline{P}^{\,* }_k$ is also negative and exponentially increasing, provided that the Euler function of degree $0$, $\overline{Q}_k$, is positive.}. All of this {\it without having made any inference or assumption} about the behavior of economic agents, except that they randomly exchange merchandise among themselves.

\subsection*{Price and Value}
\cite{Chen1}, 
in his {\it The entropy theory of value},
he made a similar proposal to ours, but defining the {\it value} as $\mathcal V = \,-\,\log_b p_k$, relating it to probability $p_k$, which he called ``scarcity". That is, the greater the ``scarcity"\ (that is, the lower the probability), the greater the value $\mathcal V$. In such a way that the value $\mathcal V(Q)$
of a certain merchandise, is not more than the average of the
value associated with each probability. Concluding then Chen that
\begin{equation}
\mathcal V(Q)= -\,\sum_{k=1}^m\,p_k\,log_b\, p_k\,.
\label{380}
\end{equation}

This proposal was made by Chen, observing exclusively the properties that $\mathcal V$ had to fulfill: a) The value of two products must be higher than the value of one of them. b) If two products are independent, that is, if the products are not substitutes or partially substitutes for each other, then
the total value of the two products must be the sum of the two products. c) The value of any product is nonnegative.

The only mathematical functions $f(x)$ that satisfy the three properties mentioned above \cite{Applebaum} are of the form $f(x) = - \log_b x$. Note that this function $f(x)$ is, except for the base of the logarithm, Eq. \ref{350} for the Euler function $\overline {S}^{\, *}_k$. The equivalent of the value function $\mathcal V(Q)$ would be the statistical average of $\overline {S}^{\,*}_k$, that is, the entropy $\overline {S}^{\,*}$.

\section{Virtual dynamics}

Through a computer program developed by us, we simulate this economic system of $N$ agents who randomly exchange $Q$ goods, and which is equivalent to the buying-selling process, where all $N$ agents can be both buyers as sellers. In each step of the simulation, the only thing that the system registers is the number of goods that each agent has.

For practical reasons, we set a {\it unit of measure} $\delta$, that is, the minimum amount of merchandise, other than zero, that each agent can trade. Here we chose as the unit of measure $\delta = 1/20$. Thus we define $\overline{Q} = \mathbf{\overline {Q}}/\delta$ as the {\it Commodity Unit}. Each of the $N$ agents is assigned at the beginning of the dynamic the same amount of merchandise, $Q/N$. In the merchandise exchange process, that is, in the buying-selling process, at any step $t_2$, two randomly selected agents $i$ and $j$ exchange goods according to the following equations
\begin{equation}
\overline{Q}_i (t_2)= \mathsf{a}\,(\overline{Q}_i(t_1)+\overline{Q}_j(t_1))\,,
\label{32}
\end{equation}
\begin{equation}
\overline{Q}_j (t_2)= (1-\mathsf{a})\,(\overline{Q}_i(t_1)+\overline{Q}_j(t_1))\,.
\label{34}
\end{equation}
Where $\overline{Q}_l (t_2)$ and $\overline{Q}_l (t_2)$ are the goods bought or sold by agent $l$ in step $t_2$; and $\overline{Q}_l (t_1)$ and $\overline{Q}_l (t_1)$ are the goods bought or sold by agent $l$ in step $t_1$, respectively. $t_2\, >\,t_1$ and $l=1$ or $2$. $\mathsf{a}$ is a random number between $0$ and $1$. In this dynamic we will call a {\it step} when the pair $i$,$j$ is chosen at random and they exchange goods, that is, they buy-sell, according to the two previous equations. A {\it run} is completed when $10$ million steps are totaled. Each result reported here is the average of $30$ runs, unless otherwise noted. From the dynamics, the number of agents $n_k$ who own a quantity of merchandise $\overline{Q}_k$ is statistically determined.

The expressions \ref{32} and \ref{34} ensure that the goods sold for $i$ are exactly the same as those obtained for $j$; in order to keep constant the total quantity of goods within the system. It is clear that in any step of the dynamic, $i$ is the {\it seller} and $j$ is the {\it buyer}. But since agents $i$ and $j$ are chosen at random in each step, $i$ could be the {\it buyer} and $j$ the {\it seller} in any other step. We chose to work with $N=3000$ agents, which is an amount that statistically represents reasonably well systems with many more agents. The maximum amount of goods that any agent can access (``the roof"\ ), $\mathbf{\overline {Q}}_m = \mathbf{Q}_m / N$, we made it vary between $10$ ({\it Upper Limit } very high) and $1.1$ ({\it Upper Limit } very low).

After obtaining the probabilities $p_k$, for each set ($\overline{Q}$, $\overline{Q}_m)$, $\overline{S}^{\,*}_k$ was graphed as a function of $\overline{Q}_k$, such that {\it assuming linear behavior} for all $k$
\footnote{Straight line $y(x)= m_p\,x + b_o$, where $m_p = (y(x_2) - y(x_1))/(x_2 - x_1)$ is the slope and $b_o$ is the cutoff with $y(x)$.},
you have
\begin{equation}
\overline {S}^{\,*}_k \cong  b\,\overline{Q}_k - \overline {A}^{\,*}_k
\,,
\label{400}
\end{equation}
thus allowing to determine, simultaneously, the slope of the straight line, the unit price $b$, and the cut with the axis of the ordinates, $\overline {A}^{\,*}_k$.

\begin{figure}[t]
	\centering
  \includegraphics[width=12cm]
    {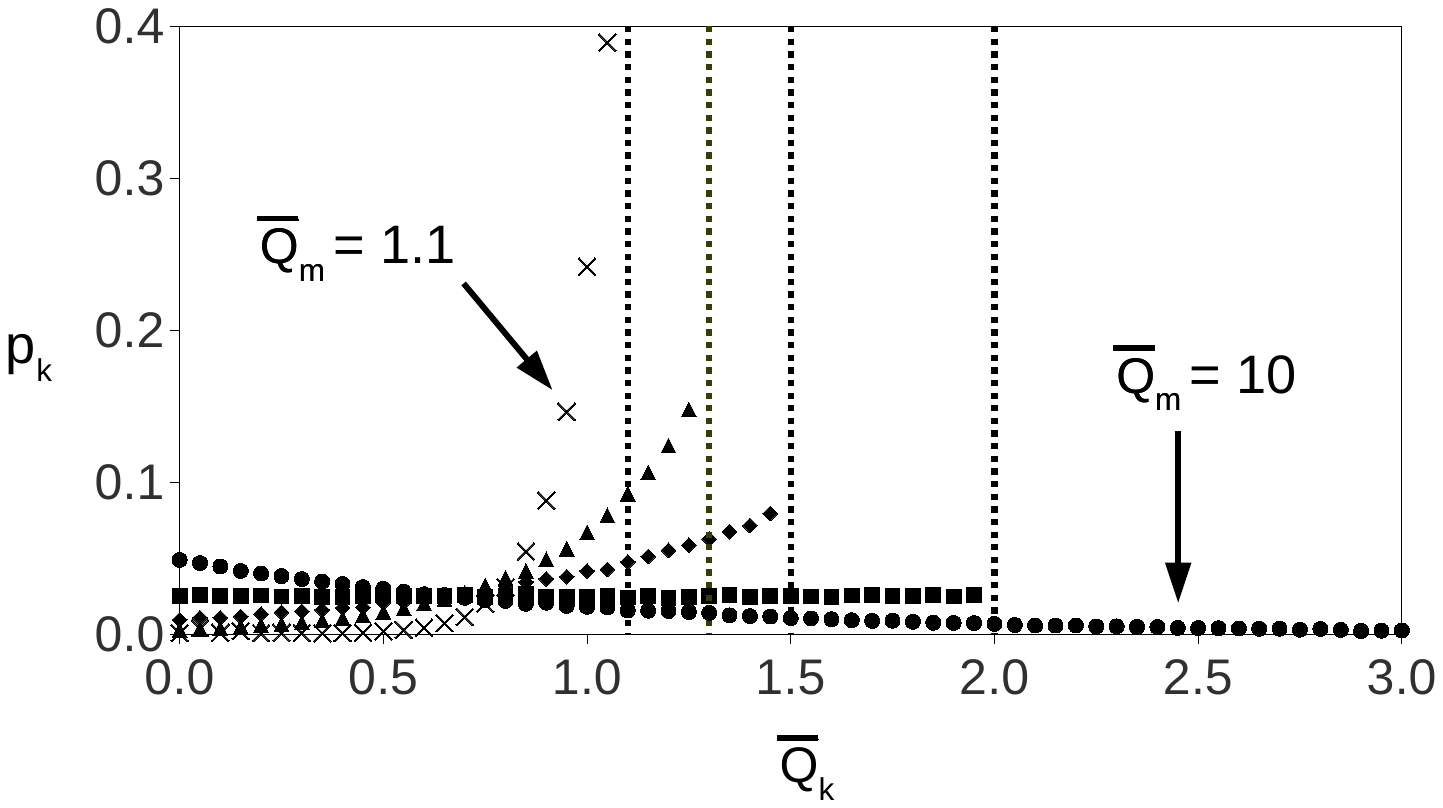}
 \caption{Probability $p_k$ vs $\overline{Q}_k$ , for different values of the "roof" $\overline{Q}_m =\,$ $10$$(\bullet)$, $2$$(\blacksquare)$, $1.5$$(\blacklozenge)$, $1.3$$(\blacktriangle)$, and $1.1$ $(\times)$.
 \label{P-VS-UG102151311}}
\end{figure}


\section{Analysis of results}


\noindent

Without loss of generality, the goods and/or commodities that are exchanged can be money itself. That money, in turn, could be the product of processes of buying and selling merchandise other than money; or the money itself (measured in $um$) involved in the purchase of foreign currency (measured in $um_0$).

\begin{figure}[ht]
	\centering
 \includegraphics[width=12cm]
    {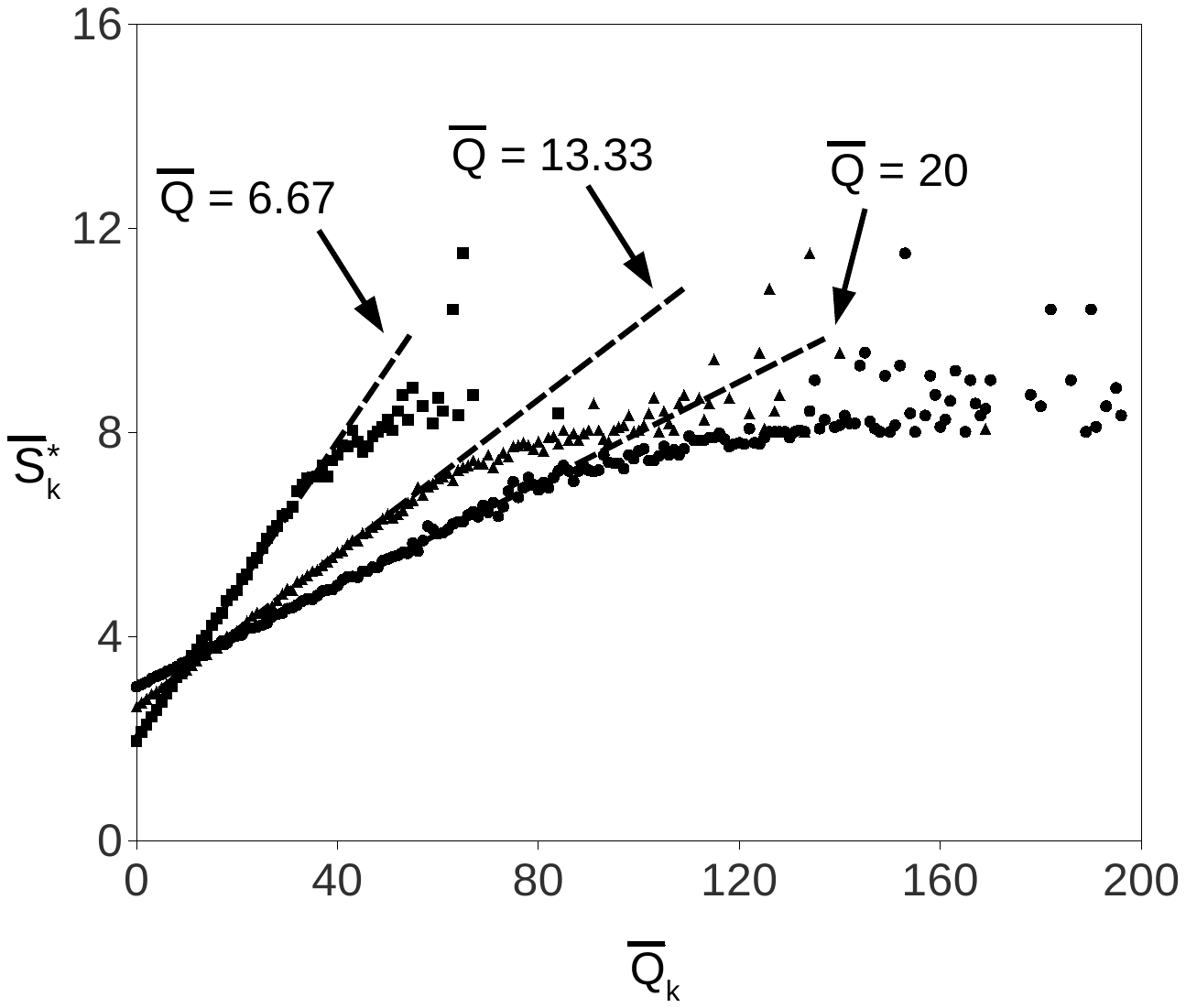}
 \caption{Euler entropy $\overline {S}^{\,*}_k$ vs $\overline{Q}_k$, from top to bottom, for three values of $\overline{Q}$ = $6.67 (\blacksquare), 13.33 (\blacktriangle)$ and $20 (\bullet)$, and $\overline{Q}_m = 10$.
  The dashed lines ($-\, -\, -\, -\,-$) represent the linear behavior of $\overline {S}^{\,*}_k$, Eq. \ref{400}. \label{EE-VS-UGK123}
 }
\end{figure}

\begin{figure}
	\centering
 \includegraphics[width=12cm]
    {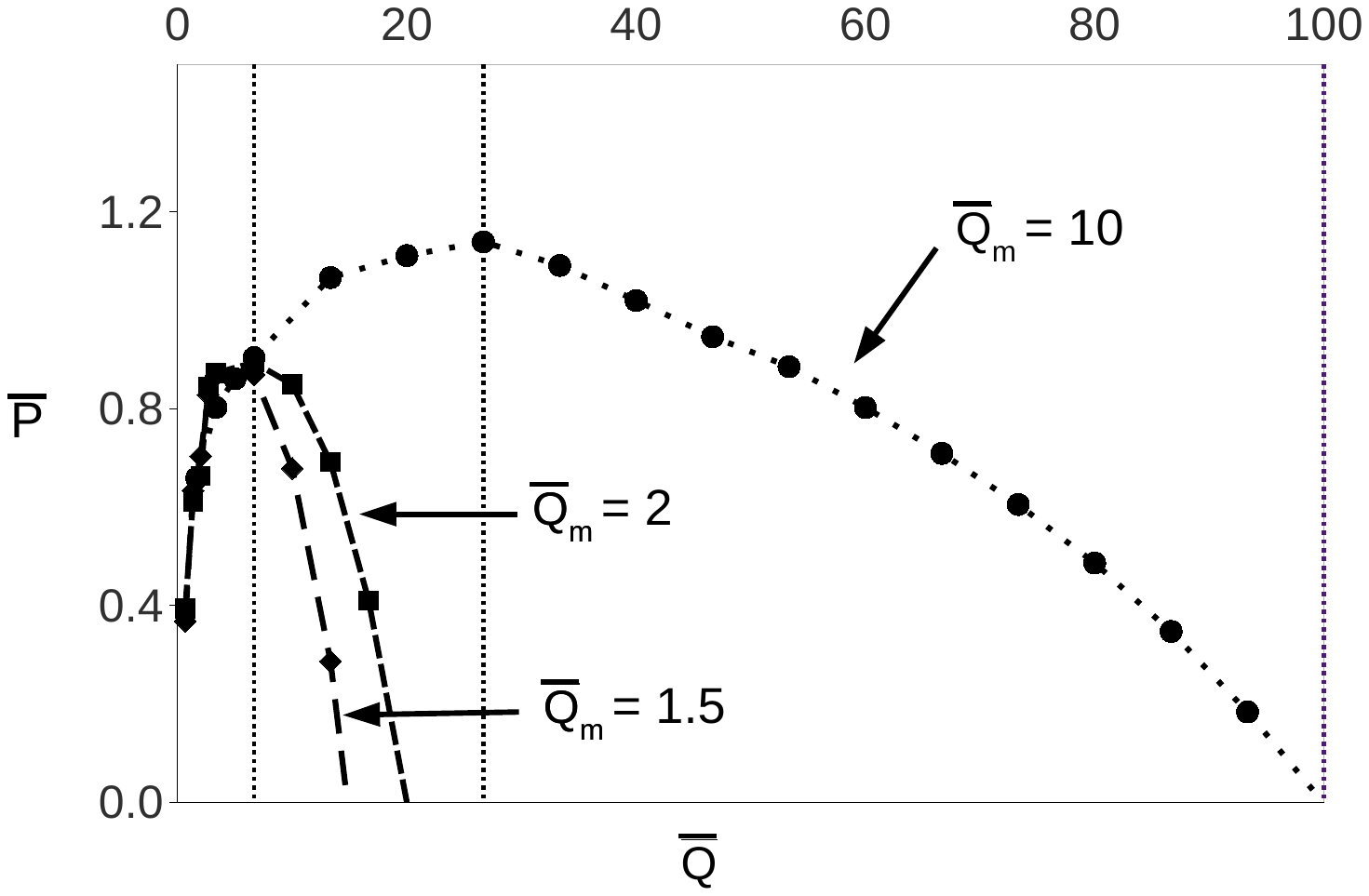}
 \caption{Total Revenue $\overline{P}$ vs Unit of Goods $\overline{Q}$, for positive values of $\overline{P}$, and roofs $\overline{Q}_m = 10$  $(\bullet)$, $2$ $(\blacksquare)$ and $1.5$ $(\blacklozenge)$. The segmented and dotted lines are a guide for the eyes. The vertical lines indicate the maximum $\overline{Q}_{MAX}$.
 \label{TR-VS-UG10215}}
\end{figure}

\begin{figure}
	\centering
 \includegraphics[width=12cm]
    {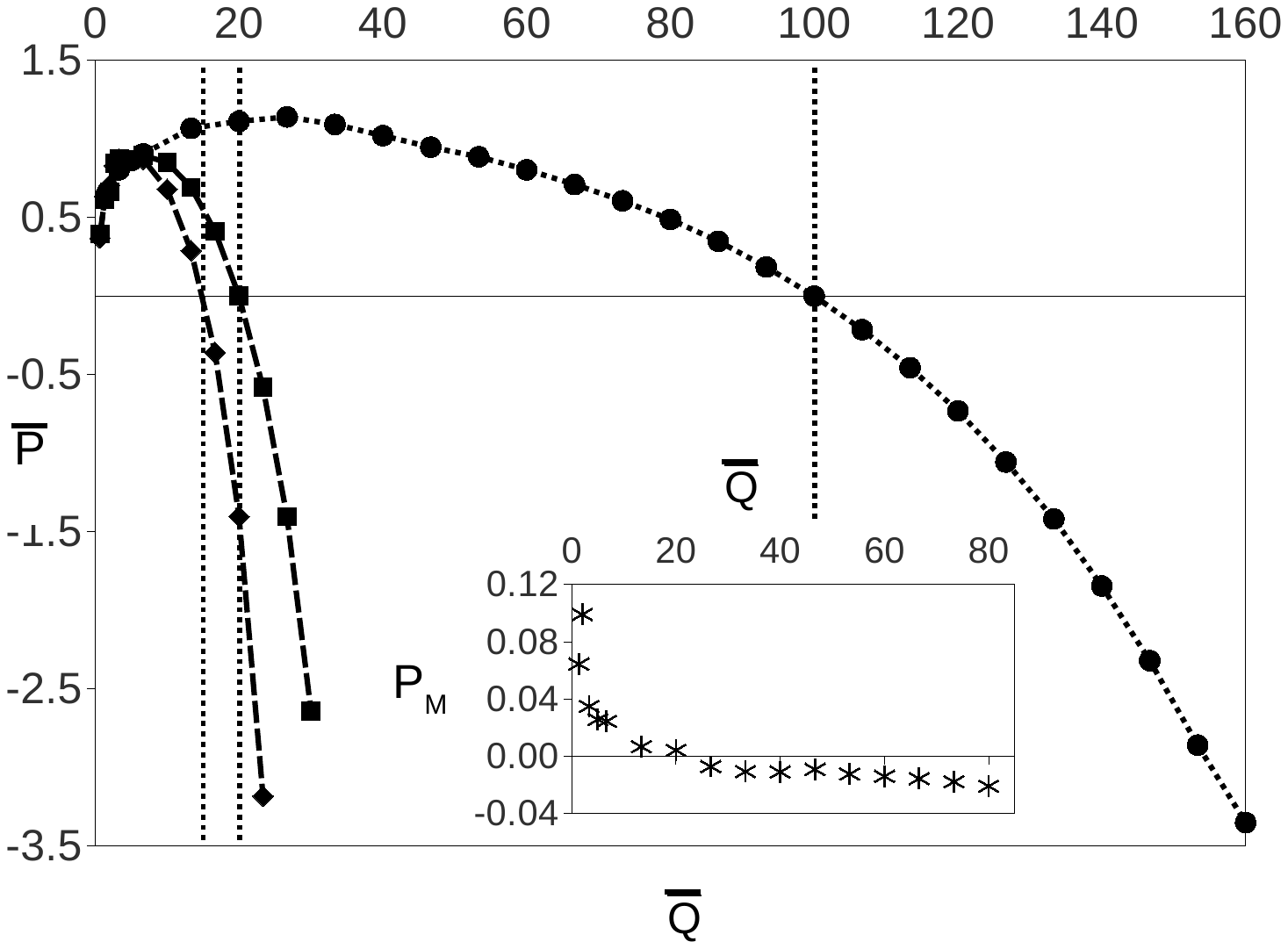}
 \caption{Total Revenue $\overline{P}$ vs Unit of Goods $\overline{Q}$, for $\overline{Q}_m = 10$ $(\bullet)$, $2$ $(\blacksquare)$ and $1.5$ $(\blacklozenge)$. The segmented and dotted lines are a guide for the eyes. The vertical lines indicate where $\overline{P}$ changes sign. The inset figure shows Marginal Revenue $P_M$ as a function of $\overline{Q}$ for $\overline{Q}_m = 10$.
  \label{TR-VS-UG10215total} }

\end{figure}

\begin{figure}
	\centering
 \includegraphics[width=12cm]
    {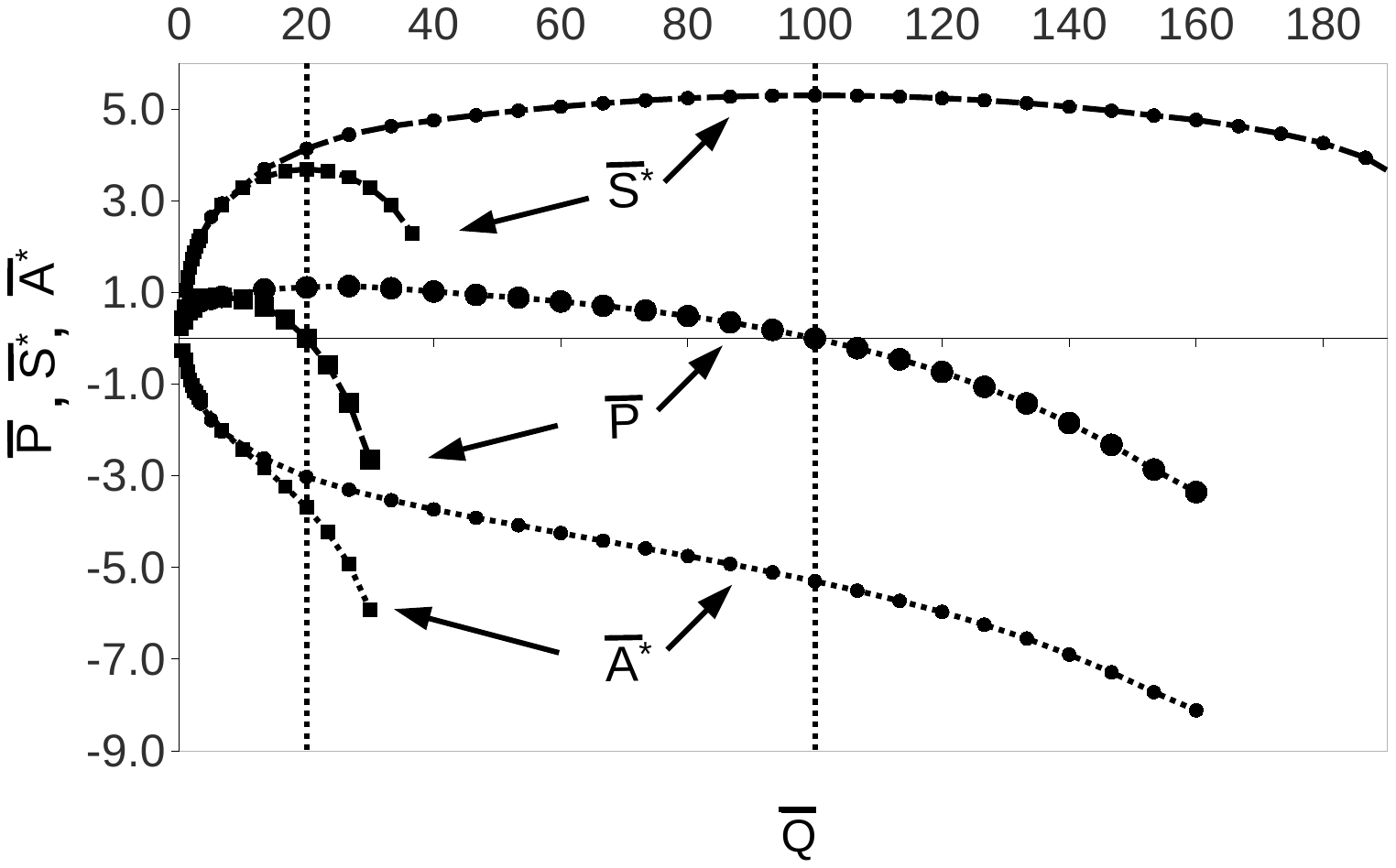}
 \caption{ \label{TREEFE-VS-UG} Total Revenue $\overline{P}$, Entropy $\overline {S}^{\,*}$, and $\overline {A}^{\,*}$, as a function of $\overline{Q}$, for roofs $\overline{Q}_m =\,$ $10$ $(\bullet)$ and $2$ 
 $(\blacksquare)$. 
   }
\end{figure}

\begin{figure}
	\centering
 \includegraphics[width=12cm]
    {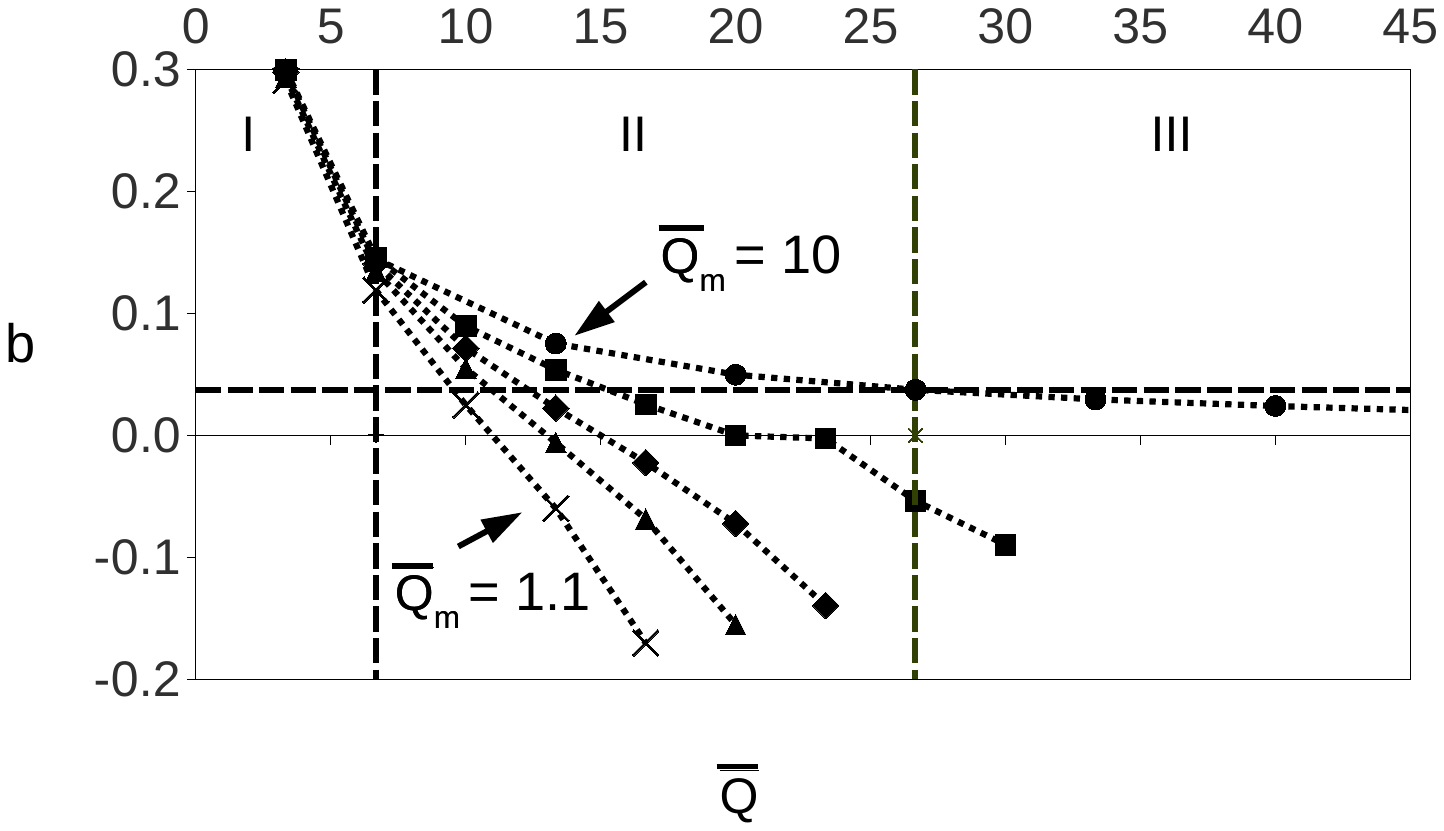}
 \caption{\label{PRICEB-VS-UG} The unit price $b$ as a function of $\overline{Q}$, for different values of $\overline{Q}_m =\,$ $10$ $(\bullet)$, $2$$(\blacksquare)$, $1.5$$(\blacklozenge)$, $1.3$$(\blacktriangle)$, and $1.1$ $(\times)$.
  }
\end{figure}

\begin{figure}
	\centering
 \includegraphics[width=12 cm]
    {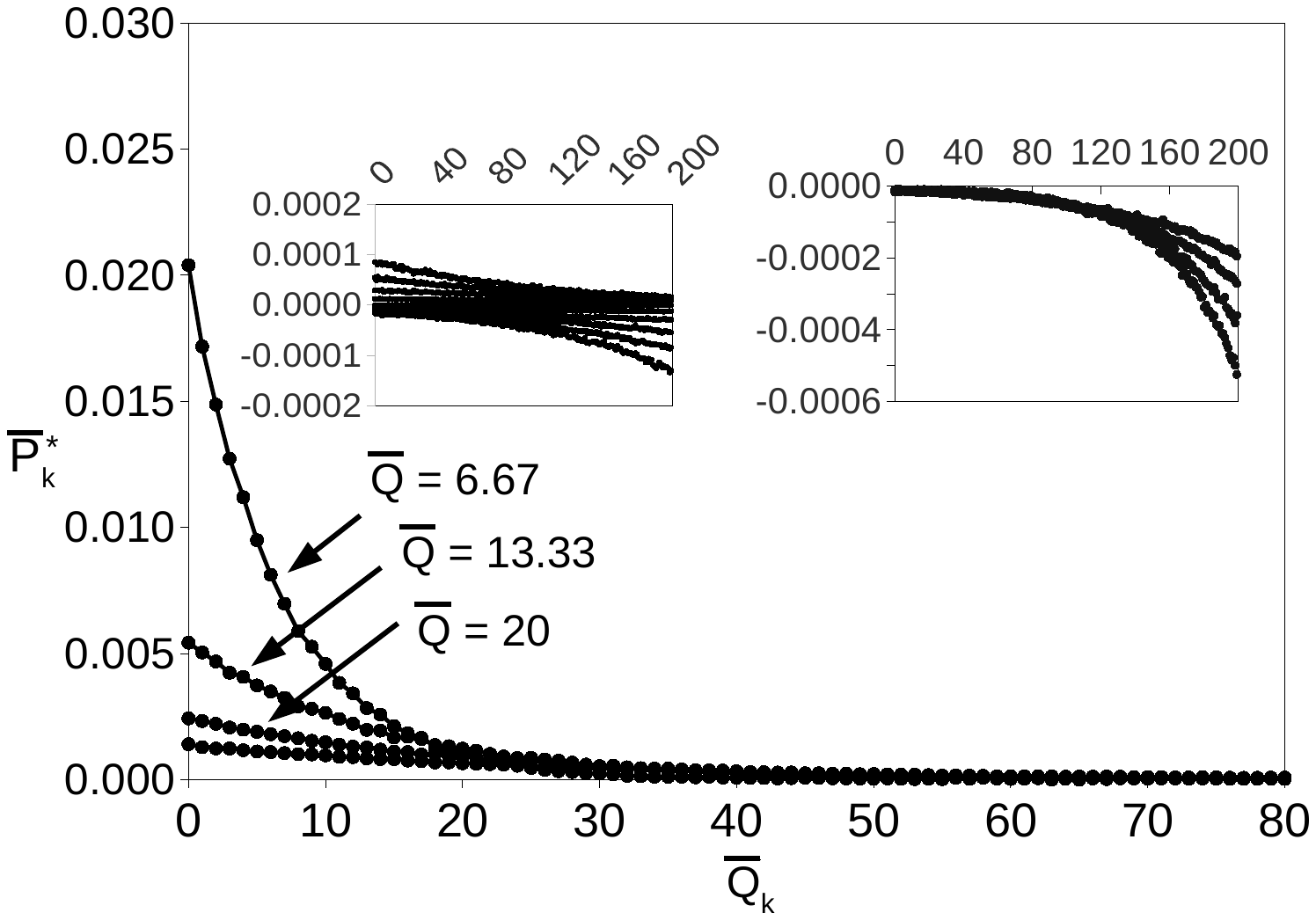}
 \caption{\label{PRICEK-VS-UGK10} Price $\overline {P}^{\,*}_k$ as a function of $\overline{Q}_k$, for 4 values of $\overline{Q}$ = $6.67$ up to $26.67$. The two inset figures show $\overline {P}^{\,*}_k$ vs $\overline{Q}_k$, for $\overline{Q}$ = $73.33$ to $133.33$ (left figure), and $ \overline{Q}$ = $140$ up to $160$ (right figure). For all of them, $\overline{Q}_m =\,$ $10$. From top to bottom, in all the figures, the value of $\overline{Q}$ increases.
     }
\end{figure}

Fig. ~\ref{P-VS-UG102151311} shows the probability profile $p_k$ as a function of the Goods Unit $\overline{Q}_k$, with $\overline{Q} = 20$
\footnote{In the work of Villagómez-Manrique et al, $\mathbf{\overline {Q}} = 1$.},
for different values of the "ceiling" $\overline{Q}_m = 10,\, 2,\, 1.5,\, 1.3\,$ and $\,1.1$, in an economy of agents that exchange money, of according to the dynamics \ref{32} and \ref{34}. For all purposes, a value of $\overline{Q}_m = 10$ will be considered to be large enough to be considered $\,\,$``without roof"\ .

The results presented here exactly reproduce those of Fig. 1 of the work by 
\cite{VillaGomez}. The most striking of these probability profiles is the {\it population inversion} that is observed when $\overline{Q}_m$ is reduced to values less than $2$, making it more likely to observe more agents with large amounts of money and very few with little money. For this to happen under these conditions, the price $b$ must be negative. Villagómez-Manrique et al interpret it by saying that the absolute temperature $T$ is negative. The equivalence between the equations \ref{365} and \ref{420} is evident.

As shown by the method used to compute $b$ and $\overline {A}^{\,*}_k$, the Euler Entropy $\overline {S}^{\,*}_k$ vs $\overline{Q}_k$, according to Eq. \ref{400}, for three values of $\overline{Q} = 6.67,\, 13.33,\, 20$, and a roof $\overline{Q}_m = $10, 
Fig. ~\ref{EE-VS-UGK123}. If $\overline {S}^{\,*}_k$ depended linearly on $\overline{Q}_k$, over the entire range of values $\overline{Q}_k$, then the result would be the straight line shown . It is clear that the experimental results differ appreciably from this behavior, for high values of merchandise $\overline{Q}_k$, even when the vast majority of agents maintain low quantities of merchandise $\overline{Q}_k$ for positive unit prices. $b$ and $\overline {A}^{\,*}_k$ were determined, respectively, from the slope and the intersection with the ordinates of the straight lines.

The Total Revenue $\overline{P}$ when buying-selling $\overline{Q}$ units of goods, for three cases, when $\overline{Q}_m = $10, $2$ and $1.5$, is shown in Fig. ~\ref{TR-VS-UG10215}. The general shape of the $\overline{P}$ corresponds to what is expected, Fig. ~\ref{TRMRLIBRO-VS-Q}. That is, as more goods are exchanged, the more the seller gets, and the buyer pays, even though the unit price $b$ will be smaller and smaller. Eventually $ \overline P$ reaches a maximum, when the decrease in $b$ is balanced with the increase in $\overline{Q}$, after which, even when more merchandise is traded, since the unit price is increasingly smaller, the total quantity of money begins to decrease. The vertical lines show the commodity values where the maximum values of $\overline{P}$, $\overline{Q}_{MAX}$ are obtained
\footnote{For the case of $\overline{Q}_m = 2$ and $1.5$, the maximum of the two coincide approximately, $\overline{Q}_{MAX} \cong 6.66$.}.
This general behavior occurs for very high roofs, like $\overline{Q}_m = 10$ ($\overline{Q}_{MAX} \cong 26.66$), and those with much lower roofs, like $\overline{ Q}_m = $2 and $1.5$. For smaller values, not shown here, the general behavior described above is still observed.

In the real world, the decrease of $\overline{P}$, after reaching the maximum, is rarely observed
\footnote{With the notable exception of the primary sector (raw materials / natural resources: agriculture, mining, etc.).}, since the economic process ends up being unfeasible for the seller, and consequently it is immediately stopped.

The complete graph of the behavior of $\overline{P}$ as a function of $\overline{Q}$ is shown in Fig. ~\ref{TR-VS-UG10215total}. The vertical lines indicate when Total Revenue $\overline{P}$ changes sign. In all the conditions studied this is observed, but at very low roofs this change in sign appears at very low commodity values. If the roof is high enough, as is the case with $\overline{Q}_m = 10$, the change in sign occurs at a fairly high value, $\overline{Q} \cong 100$. Negative $\overline{P}$ values necessarily imply that the unit price $b$ is also negative. For $\overline{Q} = 2$ and $1.5$, $\overline{P}$ changes sign by approximately $20$ and $15$, respectively.

If we now calculate the Marginal Revenue $P_M$ as the change in $\overline{P}$ due to the increase in the quantity of goods $\overline{Q}$, $P_M = \Delta {P}/ \Delta {Q } = \Delta \overline{P}/ \Delta \overline{Q}$, for the case with roof $\overline{Q}_m = 10$, the result is shown in the inset of Fig. ~\ref{TR-VS-UG10215total}. Qualitatively, it is what we would expect, Fig. \ref{TRMRLIBRO-VS-Q}, 
given the form of Total Revenue, that is, as more merchandise is exchanged, the growth of what the seller obtains "slows down" to the same extent, until it reaches zero, and then begins to be negative, at which point the maximum of $\overline{P}$ is exceeded, which represents, precisely, the maximum profit that the seller obtains.

Total Revenue $\overline P\,= \,\overline {S}^{\,*} + \overline {A}^{\,*}$ has two components, one entropic $\overline{S}^{ \,*}$ and another ``energetic"\ $\overline{A}^{\,*}$. In Fig. \ref{TREEFE-VS-UG} these three quantities are shown, for $\overline{ Q}_m = 10$ and $2$. In both cases, the maximum of entropy $\overline{S}^{\,*}$ and the change in concavity of $\overline{A}^{\,* }$, coincide with the sign change in $\overline{P}$, and therefore also in the unit price $b$.The increase in $\overline P$ (that is, when $P_M > 0$) is the commitment of two tendencies: the increase of the entropy of the system and the decrease of the factor $\overline{A}^{\,*}$. 
When $(\partial \overline{P} / \partial \overline{Q}) = 0$, then at the maximum $(\partial \overline {S}^{\,*} / \partial \overline{Q}) = - (\,\partial \overline {A}^{\,*} / \partial\overline{Q})$. Thereafter, the decrease in $\overline{P}$ will be the consequence of a greater decrease in $\overline {A}^{\,*}$ than the increase in $\overline {S}^{\,* }$. As expected, the maximum of the entropy $\overline {S}^{\,*} $ coincides with the point where $b$ and $\overline{P}$ change sign.

A crucial matter that we have investigated is the appearance of the price $b$ as a {\it natural consequence of the way} in which the merchandise is distributed. Fig. ~\ref{PRICEB-VS-UG} shows the monotonic decrease of $b$ as $\overline{Q}$ increases, for four roofs $\overline{Q}_m = 10, 2, 1.5, 1.3$ and $1.1$. That is, as more merchandise is traded, a {\it natural} decrease in the observed price is observed. In other words, the behavior observed in real economic systems is evident here, where the unit price decreases due to the increase in the quantity of goods committed in the purchase-sale process. This is well-known {\it Law of demand}. It is the result of the phenomenological observation that "when the price of a good goes up, the quantity demanded goes down", the product of certain fundamental assumptions of human behavior. For this reason, economists "believe" that this is always, or almost always, true 
\cite{Landsburg6}.

The so-called {\it retail} price, that is, when buying-selling small quantities, is statistically always higher than when buying-selling {\it wholesale}. All this without individual agents acting rationally, looking for some kind of benefit. Another way of interpreting it is that the more merchandise there is in the system, the lower the observed unit price $b$ will be, and vice versa.

Keep in mind that in this work the independent variable is the amount of goods $\overline{Q}$ or $\overline{Q}_k$, and that due to the distribution of these among the agents, the price $b$ is generated {\it naturally}.
From the economy, on the contrary, given a unit price $b$ present in the economic system, this determines the corresponding amount of merchandise traded $\overline{Q}$ or $\overline{Q}_k$ by the agents .

On the other hand, note that at $\overline{Q} \cong 100$, $20$, and $5$, there is a sign change in the respective unit price $b$, as expected. Likewise, in the real world of economics, it is extremely rare to see negative prices, and this has value here only for academic reasons. Clearly, in this case, the density profiles $p_k$ must also present population inversion.

An advantage of the methodology described here is that it is possible to answer, in a simple way, the reasons for a basic phenomenon observed in the economy: The depreciation of the country's currency due to the issuance of money ``created"\ by the Central Bank
\footnote{We abstract from the secondary expansion of the banks, as a consequence of the granting of credits, as another source of ``creation"\ of money.},
how much it injects money, so that the government, for example, can meet salary commitments with its workers.

In our economic system model, whatever the roof is set, the price $b$ 
\footnote{Remember that $b$ is a relative value of $um/um_0$.}
decreases as the quantity of goods $\overline{Q}$ increases. If the commodity is money, then money itself it {\it depreciates} when more money is in the hands of the economic agents. However, the higher the roof $\overline{Q}_m$, the more ``slowly"\ will be its depreciation, Fig.~ \ref{PRICEB-VS-UG}. Below $\overline{Q} \cong 6.66$, indicated by a vertical line, the behavior of the system is
essentially the same, regardless of the ceiling, noted in figure as region I. Between that line and the second line located in
$\overline{Q} \cong 26.66$, region II, Total Revenue $\overline{P}$ is in the decreasing zone for all but the highest roof
$\overline{Q}_m =10$. From a practical point of view, this area is you'll see little in the real world for low roofs. Likewise, for
above this line, region III, $\overline{P}$ is also decreasing for $\overline{Q}_m =10$.

A decreasing $\overline{P}$, that is, a Marginal Revenue $P_M < 0$, is not convenient for whoever ``sells"\ the money (the Central Bank), since it receives less currency from the government for more money created. Here we have considered the case when the Central Bank ``sells"\ money ($um$) to the government, who ``pays"\ it with foreign currency ($um_0$) product of the collection of taxes on foreign capital, or from the activity of its mixed or state companies, which sell raw materials or manufactured products outside the country. With this money, the government pays its national employees and suppliers, recovering it later through the taxes it collects. Then it returns to the Central Bank and it exchanges them for currencies, at a certain price, which will be used to purchase products abroad, pay debt, etc., thus completing the cycle.

So, one way to "slow down"\ the depreciation of money, and keep its price $b$ constant, due to the issuance of more money, is to raise the roof $Q_m$. That is, the economic space created must be ``expanded"\, so as to remain in the area where the Marginal Revenue ${P}_M$ is always positive. This is what we will call here {\it indexing}, that is, a mechanism tending to preserve the price of money, that is, its purchasing power, by increasing the flow of money in the economy. The dotted horizontal line in Fig. \ref{PRICEB-VS-UG} allows us to understand this process, focusing mainly on region II. It is clear that we must "jump"\ to higher and higher ceilings if we want to keep the price $b$ of money stable, as we increase its quantity in the economy.

An issue that may also be of interest is to estimate the value of the price at each site $k$, $\overline {P}^{\,*}_k$.  As can be seen in Fig. \ref{PRICEK-VS-UGK10}, $\overline {P}^{\,*}_k$ follows the general decreasing behavior of $b$, as the quantity of merchandise increases $\overline{Q}$.  In other words, this is the ``Law of demand"\ before described.
The sign of $\overline {P}^{\,*}_k$ also corresponds to the sign of $b$. The unit price $b$, with $\overline{Q}_m = 10$, is positive for $0\,\le \,\overline{Q}\, \leq \,100$, and negative for $\overline{Q}\, > \,100$.

\section{Conclusions}

In this work we have developed a methodology based on thermodynamic and mechanical-statistical concepts, which facilitate the conceptual understanding of phenomena that have been widely observed in real economic systems, such as the price, the Law of demand, the Total Revenue, and the depreciation of a currency. The latter occurs when the economic system is flooded with more money, and the way to index prices is by expanding the economy, that is, increasing the "roof" of the economic system. These phenomena arise as a consequence of two quantities that compete with each other: configurational entropy and a term such as "free energy", a product of the {\it way} in which goods are distributed between the economic agents. The unit price $b$, whose value {\it emerges naturally}, is equivalent to $\beta = 1/ k_B T$ in classical thermodynamic systems, presenting situations where there are negative prices. In short, using Euler's theorem of homogeneous functions has made it possible to connect economic properties and phenomena with their thermodynamic equivalents.

\section*{Acknowledgments}

We would like to thank Marcelo del Castillo-Mussot for his valuable opinions and recommendations.

%
\bibliographystyle{unsrtnat}

\end{document}